\documentclass[preprint,prd,nofootinbib,tightenlines,amsmath]{revtex4}
\usepackage{graphicx}% Include figure files
\usepackage{bm}% bold math
\usepackage{color}
\usepackage{subfigure}

\begin{document}
\baselineskip=15pt \parskip=5pt

\vspace*{3em}

\title{Low Mass Dark Matter and Invisible Higgs Width \\In Darkon  Models}

\author{Yi Cai$^1$}
\author{Xiao-Gang He$^{1,2}$}
\author{Bo Ren$^1$}

\affiliation
{$^1$INPAC, Department of Physics, Shanghai Jiao Tong University, Shanghai, China\\
$^2$Department of Physics and  Center for Theoretical Sciences, \\
National Taiwan University, Taipei, Taiwan}

\date{\today $\vphantom{\bigg|_{\bigg|}^|}$}

\begin{abstract}
The Standard Model (SM) plus a real gauge-singlet scalar field dubbed darkon (SM+D) is the
simplest model possessing a weakly interacting massive particle (WIMP) dark-matter candidate.
In this model, the parameters are constrained from dark matter relic density and direct searches.
The fact that interaction between darkon and SM particles is only mediated by Higgs boson exchange may lead to
significant modifications to the Higgs boson properties.
 If the dark matter mass is smaller than a half of the Higgs boson mass,
the Higgs boson can decay into a pair of darkons
resulting in a large invisible branching ratio.
The Higgs boson will be searched for at the LHC and may well be discovered in the near future.
If a Higgs boson with a small invisible decay width will be found,
 the SM+D model with small dark matter mass will be in trouble.
We find that by extending the SM+D to a two-Higgs-doublet model plus a darkon (THDM+D) it is
possible to have a Higgs boson with a small invisible branching ratio and at the
same time the dark matter can have a low mass.
We also comment on other implications of this model.
\end{abstract}

\maketitle

\section{Introduction}

Various astronomical and cosmological observations show that there is dark matter
 (DM) making up about 20\% of the energy of our universe.
 Although the evidence for DM has been established for many decades,
the identity of its basic constituents has so far remained elusive.
One of the popular candidates for DM is the weakly interacting massive particle (WIMP).
Among a large number of possible WIMPs,
the lightest supersymmetric particle has been studied most.
Although this possibility has many attractive features,
direct experimental evidence for it has not been discovered.
There are other possible WIMPs which can explain the DM relic density.
The simplest model is the SM+D, which extends the SM by the addition
of a real gauge-singlet scalar field $D$.
This singlet field, the darkon field, can play the role of DM.
The darkon as DM was first considered by Silveira and Zee~\cite{Silveira:1985rk}.
Implications of this model and some variations of it have been explored by other
groups~\cite{McDonald:1993ex,Burgess:2000yq,darkon,Bird:2004ts,Tsai,Cynolter:2004cq,pt,barger,sm+d}.
In the SM+D model, interactions between darkon and SM particles are only through
 Higgs boson exchange. This may lead to significant modifications to the Higgs boson property.

At present the DM mass is not known.
It can be as heavy as a few hundred GeV to a few TeV from indirect DM search~\cite{deBoer:2008iu},
such as PAMELA~\cite{pamela}, ATIC~\cite{atic} and FERMI-LAT~\cite{fermi-lat}. A DM of mass 50 to 70 GeV may
also explain the gamma-ray excess observed in the EGRET data~\cite{deBoer:2005tm}.
There are also evidences, such as DAMA~\cite{Savage} and CoGeNT~\cite{Aalseth},
indicating that the DM mass can be as low as a few GeV.
The implications with a light or heavy DM can be very different.
If the DM mass is smaller than a half of the Higgs boson mass,
in the SM+D model the Higgs boson can decay into a pair of DM fields,
which results in a large invisible branching ratio for the Higgs boson.
The Higgs boson will be searched for at the LHC and may well be discovered in the near future.
If a Higgs boson with a small invisible decay width will be found,
the SM+D model with small DM mass will be in trouble.
Other direct search of DM also provide constraints on the DM mass in this model
~\cite{Ahmed:2008eu,Angle:2007uj,henry-wong}.
In this work we study implications of low DM mass
on SM+D and a two-Higgs-doublet extension (THDM+D).
We find that in THDM+D it is possible to have a Higgs boson with a small invisible branching
ratio and at the same time the DM can have a low mass in the a few GeV to about a half of Higgs boson mass range.

Before discussing THDM+D, let us briefly summarize the main results
of the DM in the SM+D model.
Since the darkon $D$ must interact weakly with the SM matter fields
and be stable to play the role of DM,
the simplest way to introduce the darkon is to impose a discrete $Z_2$ symmetry
so that it can only be created or annihilated in pairs.
In a renormalizable theory $D$ can only couple to the Higgs doublet field $H$.
So besides the kinetic energy term \,$\frac{1}{2}\partial^\mu D\,\partial_\mu^{}D$,\,
the Lagrangian of the darkon interaction part takes the form~\cite{Silveira:1985rk,Burgess:2000yq}
\begin{eqnarray}  \label{DH}
{\cal L}_D^{} \,\,=\,\, -\frac{\lambda_D^{}}{4}\,D^4
- \frac{m_0^2}{2}\,D^2 - \lambda\, D^2\,H^\dagger H \,\,,
\end{eqnarray}
where  $\lambda_D^{}$,  $m_0^{}$, and $\lambda$  are free parameters.
${\cal L}_D^{}$ is invariant under the $Z_2$ symmetry where
only $D$ is odd and all other SM fields are even.
The parameters in the potential should be carefully chosen
such that $D$ does not develop a vacuum expectation value (vev) and the $Z_2$ symmetry is unbroken,
which will ensure that the darkon does not mix with the Higgs field
and thus possible fast decays into other SM particles will be avoided.

The Lagrangian in Eq.~(\ref{DH}) can be rewritten to describe the interaction of the physical
Higgs boson $h$ with the darkon as
\begin{eqnarray}
{\cal L}_D^{} \,\,=\,\, -\frac{\lambda_D^{}}{4}\,D^4-\frac{\bigl(m_0^2+\lambda v^2\bigr)}{2}\,D^2
- \frac{\lambda}{2}\, D^2\, h^2 - \lambda v\, D^2\, h \,\,,
\end{eqnarray}
where  \,$v=246$\,GeV\,  is the vev of the neutral component of $H$,
the second term contains the darkon mass
\,$m_D^{}=\bigl(m^2_0+\lambda v^2\bigr)^{1/2}$,\, and the last term,  \,$-\lambda v D^2 h$,\,
plays an important role in determining the relic density of the DM.
At the leading order, the relic density of the darkon results from the annihilation of a darkon
pair into SM particles through Higgs exchange, namely
\,$DD\to h^*\to X$,\, where $X$ indicates SM particles.

Since the darkon is cold DM, the speed of darkon is highly non-relativistic and
the invariant mass of a darkon pair is  roughly $\sqrt s\simeq2m_D^{}$.
Given the determined SM+D Lagrangian, the $h$-mediated
annihilation cross-section of a darkon pair into SM particles is then given by~\cite{Burgess:2000yq}
\begin{eqnarray} \label{csan}
\sigma_{\rm ann}^{}\, v_{\rm rel}^{} \,\,=\,\,
\frac{8\lambda^2 v^2}{\bigl(4m_D^2-m_h^2\bigr)^2+\Gamma^2_h\,m^2_h}\,
\frac{\sum_i\Gamma\bigl(\tilde h\to X_i^{}\bigr)}{2m_D^{}} \,\,,
\end{eqnarray}
where \,$v_{\rm rel}^{}=2\bigl|\bm{p}_D^{\rm cm}\bigr|/m_D^{}$\, is the relative speed of
the $DD$ pair in their center-of-mass (cm) frame, $\tilde h$  is a virtual Higgs boson
with an invariant mass $\sqrt{s}=2m_D$
which couples to the other states as the physical $h$ with mass $m_h^{}$,
and $\tilde h\to X_i$ is any possible decay mode of $\tilde h$.
For a given model, \,$\Sigma_i\Gamma\bigl(\tilde h\to X_i\bigr)$\,
is obtained by calculating the decay width of $h$ and replacing $m_h^{}$ with  $2m_D^{}$.

The elastic cross section of DM with nucleon can also be calculated
and compared with the direct DM search data.
The cross section is given by
\begin{eqnarray}
\sigma_{\rm el}^{} \,\,\simeq\,\,
\frac{\lambda^2\,g_{NN\cal H}^2\,v^2\,m_N^2}{\pi\,\bigl(m_D^{}+m_N^{}\bigr)^2\, m_{\cal H}^4} \,\,,
\end{eqnarray}
where the approximation  \,$\bigl(p_D^{}+p_N^{}\bigr)^2\simeq\bigl(m_D^{}+m_N^{}\bigr)^2$\, is used.
The Higgs-Nucleon coupling $g_{NN\cal H}$ is given by
\begin{eqnarray} \label{gnnh0}
g_{NN\cal H}^{}\, \bar N N \,\,=\,\,
\langle N| \frac{k_u^{}}{v} (m_u^{}\, \bar u u+m_c^{}\, \bar c c+m_t^{}\, \bar t t)+
\frac{k_d^{}}{v}( m_d^{}\, \bar d d+m_s^{}\, \bar s s+m_b^{}\, \bar b b )|N\rangle \,\,.
\end{eqnarray}
In the SM+D, the Higgs-nucleon coupling $g_{NNh}^{\rm SM}$ is obtained by setting
\,$k_u^{}=k_d^{}=1$ in the equation above. The Higgs-nucleon coupling has been studied in the context of the SM~\cite{Shifman:1978zn, Cheng:1988cz, gasser}.
We will use the numerical value for $g_{NNh}$ in Ref.~\cite{sm+d}
 based on chiral perturbation theory estimate,
\begin{eqnarray}
g_{NNh}^{\rm SM} \,\,\simeq\,\, 1.71\times10^{-3}  \,\,.
\end{eqnarray}

The $h\to DD$ decay width is given by
\begin{eqnarray}
\Gamma(h \to DD) = \frac{1}{8\pi}\frac{\lambda^2v^2}{m_h}\sqrt{1-\left(\frac{2 m_D}{m_h}\right)^2}.
\end{eqnarray}

\begin{figure}[htp]
\label{fig:smdlambda}\includegraphics[width=0.4\textwidth]{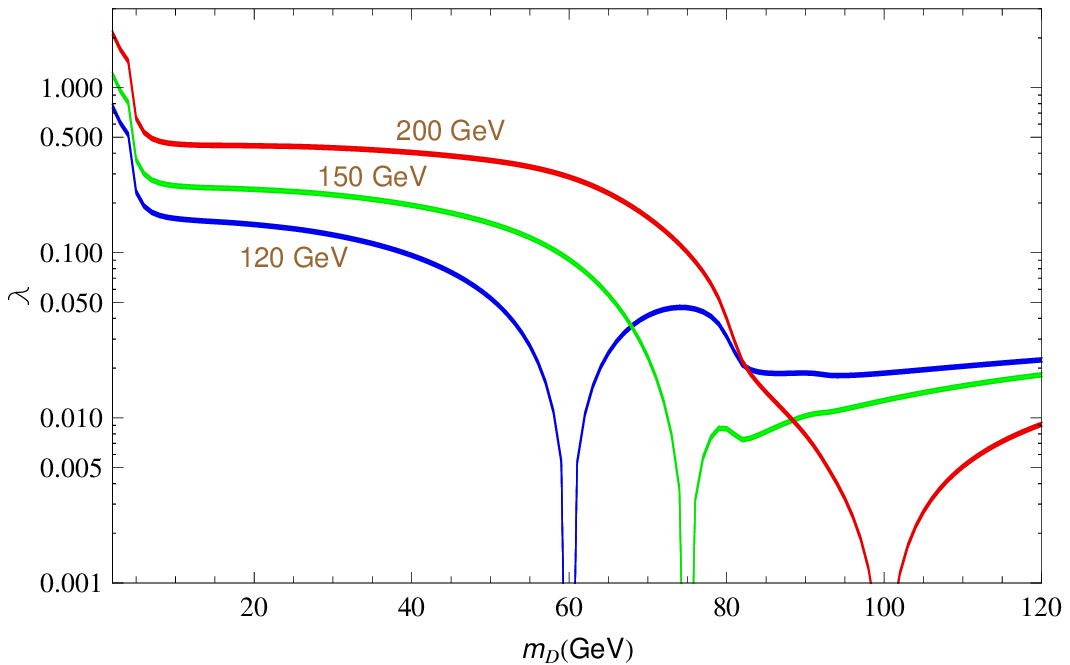}
\label{fig:smdsigma}\includegraphics[width=0.4\textwidth]{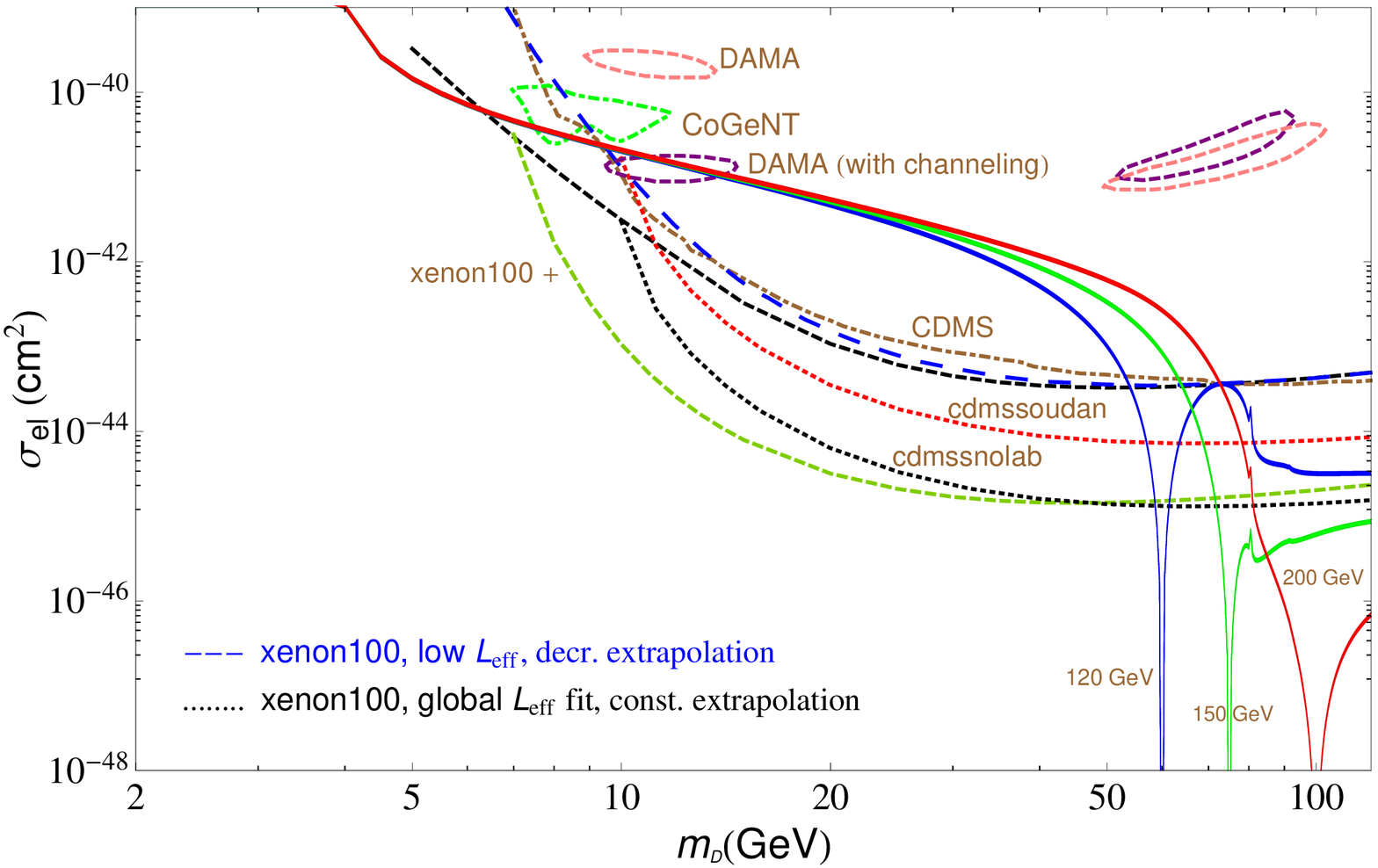}\\
\caption{
Figure on the left is for Darkon-Higgs coupling $\lambda$ as a function of darkon mass $m_D$
with different Higgs mass $m_h$ in SM+D.
Figure on the right is for Darkon-Nucleon elastic cross section $\sigma_{\rm el}$ as a function of darkon mass $m_D$
with different Higgs masses $m_h$, compared to 90\% C.L. upper
limits from DAMA~\cite{Savage}, CoGeNT~\cite{Aalseth}, CDMS~\cite{Ahmed:2008eu} and XENON~\cite{Angle:2007uj}.
Future projected experimental sensitivities for superCDMS
~\cite{superCDMS} and Xenon100+~\cite{xenon100p} are also shown.
\label{fig:smd}}
\end{figure}

There are only a small number of unknown parameters in this model.
Besides the Higgs boson mass $m_h$, the other two unknown parameters we concern are
the darkon mass $m_D$ and the coupling constant $\lambda$.
For a given $m_h$, the standard relic density calculation~\cite{dmplot,Kolb:1990vq}
can be used to constrain the allowed parameter space for $m_D$ and $\lambda$.
With this constraint, the cross section for direct DM search can be calculated
and then compared with the current direct DM searching data,
which further constrains the parameter space allowed.
Finally one can calculate the invisible branching ratio of $h \to DD$
if $m_D$ is smaller than $m_h/2$.  The predicted Higgs boson invisible
branching ratio can then be tested at the LHC.
The main results are shown in Figs.~\ref{fig:smd} and \ref{fig:smdbr}, in which the models produce
the DM relic density in the 90\% C.L. range derived from the new WMAP7 results~\cite{Nakamura}, $\Omega_D h^2 \subset [\, 0.1065, 0.1181\, ]$.

\begin{figure}[htp]
\label{fig:smdbr}\includegraphics[width=0.4\textwidth]{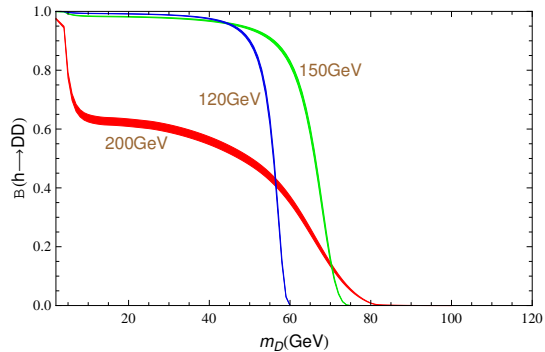}
\caption{
The invisible decay branching ratio of the Higgs boson
as a function of darkon mass $m_D$ with different Higgs boson mass $m_h$.\label{fig:smdbr}}
\end{figure}

It can be seen from Fig.~\ref{fig:smd} that a low mass DM roughly in the range of 10 GeV
 to a half of Higgs boson mass
is in conflict with direct DM search.
But the region below is still possible and may accommodate the DAMA or
the CoGeNT data.
It can also be clearly seen in Fig.~\ref{fig:smdbr} that when the DM mass is low  enough
that $h \to DD$ is kinematically allowed, the properties of the Higgs boson will be
 dramatically affected.
The Higgs boson will decay predominately into a pair of darkons, which leads to a
large invisible branching ratio as shown in Fig.~\ref{fig:smdbr}.
The invisible decay of Higgs boson can be
detected only if there are additional objects produced together with the
Higgs boson, since a hadron collider only measures the transverse missing energy.
Thus the invisible decay of the Higgs boson can be detected in the associated production
channels like $(Zh)$, $(t\bar{t}h)$ and $(q\bar{q}h)$~\cite{invisiblehiggs1, invisiblehiggs2},
which has been further studied in detail in a recent ATLAS analysis~\cite{invisiblehiggs3}.
A low DM mass will be ruled out if a Higgs boson
with a small invisible decay branching ratio is found at the LHC.

In this work we investigate whether a low mass
DM and a Higgs boson with a small invisible branching ratio
are able to be reconciled in two Higgs doublet extensions of darkon models (THDM+D).
We find that such a possibility can actually be realized in this model.\\

\section{Two-Higgs-doublet model with  a darkon}

Depending on how the two Higgs doublets $H_1$ and $H_2$ couple
to the fermions in the SM, there are three types of THDM~\cite{thdm}.
In the THDM\,I, only one of the Higgs doublets gives masses to all the fermions.
In the THDM\,II, the up-type fermions get mass from only one of the Higgs doublets, say $H_2$,
and the down-type fermions from the other doublet.
In the THDM\,III, both $H_1$ and $H_2$ give masses to all the fermions.

Since only one Higgs doublet generates the fermion masses in the THDM\,I,
the Higgs couplings to fermions are the same as in the SM, up to an overall scaling factor.
Therefore the Higgs couplings in the THDM\,I+D are similar to those in the SM+D studied in
the previous section and thus cannot help to ease the tension between the direct searches
and the invisible branching ratio.
In the THDM\,III, there are flavor-changing Higgs-quark couplings which introduce too many
parameters for the model to be predictable.
So we will concentrate on the THDM\,II with a darkon field (THDM\,II+D).

The Yukawa interactions of the Higgs fields in the THDM\,II are given by~\cite{thdm}
\begin{eqnarray} \label{yukawa_2hdm}
{\cal L}_{\rm Y}^{} \,\,=\,\, - \bar Q_L^{} \lambda^u_2 \tilde H_2^{}{\cal U}_R^{}
- \bar Q_L^{}\lambda^d_1 H_1^{} {\cal D}_R^{}
- \bar L_L^{} \lambda^l_1  H_1^{} E_R^{} \,\,+\,\, {\rm h.c.} \,\,,
\end{eqnarray}
where  $Q$, $\cal U$, $\cal D$, $L$, and $E$ represent the usual quark and lepton
fields and $\lambda^{u,d,l}$ are Yukawa couplings.
A discrete $Z'_2$ symmetry, under which $H_2$ and $U_R$ are the only odd fields,
 has to be introduced to forbid
$\bar{Q}_L \tilde{H}_1 U_R$, $\bar{Q}_LH_2 D_R$ and $\bar{L}_L H_2 E_R$ to obtain the Yukawa couplings above.
The Higgs doublets can be decomposed as
\begin{eqnarray}
H_k^{} \,\,=\,\, \frac{1}{\sqrt2} \left(\begin{array}{c} \sqrt{2} h^+_k\\
v_k^{}+ h_k^{} + i I_k^{} \end{array}\right ) \,\,,
\end{eqnarray}
where \,$k=1,2$\, and $v_k^{}$ is the vev of the neutral component of $H_k$.
Here $h^+_k$ and $I_k^{}$ are related to the physical Higgs bosons $H^+$ and $A$ and
the would-be Goldstone bosons $w$ and $z$ by
\begin{eqnarray}
\left(\begin{array}{c}h^+_1 \\ h^+_2\end{array}\right) \,\,=\,\,
\left(\begin{array}{rrr} \cos\beta && -\sin\beta \\ \sin\beta && \cos\beta \end{array}\right)
\left(\begin{array}{c}w^+\\H^+\end{array}\right) \,\,, \hspace{2em}
\left(\begin{array}{c}I_1^{} \\ I_2^{} \end{array}\right) \,\,=\,\,
\left(\begin{array}{rrr}\cos\beta &&-\sin\beta \\ \sin\beta && \cos\beta\end{array}\right)
\left(\begin{array}{c}z\\A\end{array}\right) \,\,\,,
\end{eqnarray}
with  \,$\tan\beta=v_2^{}/v_1^{}$,\, while $h_k^{}$ can be expressed in terms of mass
eigenstates $h$ and $H$ as
\begin{eqnarray}
\left(\begin{array}{c}h_1^{} \\ h_2^{} \end{array}\right) \,\,=\,\,
\left(\begin{array}{rrr} \cos\alpha && -\sin\alpha \\ \sin\alpha && \cos\alpha\end{array}\right)
\left(\begin{array}{c}h \\ H \end{array}\right) \,\,,
\end{eqnarray}
where the angle $\alpha$ is the mixing of the two $CP$-even Higgs bosons.
The would-be Goldstone bosons $z$ and $w^\pm$ will be eaten by $Z$ and $W^\pm$ respectively.
In the limit where $\alpha = \beta$,
$h$ has the same couplings to other SM particle as the Higgs boson in SD+D model. Then
$h$ is the SM-like Higgs in this sense.
There are, however, enough degrees of freedom in the Higgs potential where the mixing angle
$\alpha$ can have a large deviation from $\beta$,
and $h$ can also be either lighter or heavier than $H$.
We will consider both mass hierarchy cases in our discussion later.

Various couplings of Higgs bosons to the other SM fields and the darkon
can then be re-written in the mass eigenbasis.
In analogy to Eq.~(\ref{DH}) in the SM+D case,  in the THDM\,II+D  we have the renormalizable
darkon Lagrangian
\begin{eqnarray}
{\cal L}_D^{} \,\,=\,\, -\frac{\lambda_D^{}}{4}\, D^4 - \frac{m_0^2}{2}\, D^2
- \bigl(\lambda_1^{}\, H_1^{\dag}H_1^{} + \lambda_2^{}\, H^\dagger_2 H_2^{} \bigr) D^2 \,\,.
\end{eqnarray}
As in the SM+D, we have again imposed the $Z_2$ symmetry under which only $D$ is odd.
For the same reasons it has to be unbroken.
This Lagrangian also respects the $Z'_2$ symmetry mentioned earlier.

After electroweak symmetry breaking, the darkon Lagrangian ${\cal L}_D$ contains
the $D$ mass term and the $DD(h,H)$ terms are given by~\cite{sm+d}
\begin{eqnarray} \label{lambda}
\nonumber
m_D^2 &=& m_0^2 +(\lambda_1 \cos^2\beta + \lambda_2 \sin^2\beta)v^2,\\
{\cal L}_{DDh} &=& - (\lambda_1 \cos\alpha \cos\beta + \lambda_2 \sin\alpha \sin\beta) v D^2 h = -\lambda_h v D^2 h\\
\nonumber
{\cal L}_{DDH} &=& - (-\lambda_1 \sin\alpha\cos\beta+ \lambda_2 \cos\alpha \sin\beta)v D^2 H  = -\lambda_H v D^2 H,
\end{eqnarray}
where  $v^2=v_1^2+v_2^2$.
There is, however, no $DDA$ term in ${\cal L}_D$.
Since $m_0$, $\lambda_1$ and $\lambda_2$ are all free parameters, we can treat the darkon mass
$m_D$ and the effective couplings $\lambda_{h,H}$ as new free parameters in this model.

From Eq.~(\ref{yukawa_2hdm}), the Yukawa interactions of $h$ and $H$ are described by~\cite{thdm}
\begin{eqnarray}
{\cal L}_{ff\cal H}^{} &=& -\bar{\cal U}_{L} M^u {\cal U}_R^{}\,
\Biggl(\frac{\cos\alpha}{\sin\beta}\,\frac{H}{v}
       + \frac{\sin\alpha}{\sin\beta}\,\frac{h}{v}\Biggr)
- \bar{\cal D}_L^{} M^d {\cal D}_R^{}\, \Biggl(-\frac{\sin\alpha}{\cos\beta}\,\frac{H}{v}
+ \frac{\cos\alpha}{\cos\beta}\,\frac{h}{v}\Biggr)
\nonumber\\ &&
-\,\, \bar E_L^{} M^l E_R^{}\, \Biggl(-\frac{\sin\alpha}{\cos\beta}\,\frac{H}{v}
+ \frac{\cos\alpha}{\cos\beta}\,\frac{h}{v}\Biggr)  \,\,+\,\, {\rm h.c.}  \,\,.
\label{yukawa}
\end{eqnarray}
We have not written down the couplings of the $CP$-odd Higgs boson $A$ to fermions because it
does not couple to $D$. Consequently $A$ is irrelevant to our darkon relic-density study
and direct DM search.

We now write down explicitly the couplings between the CP-even Higgs bosons and the vector bosons
which are relevant to the Higgs decay calculation.  We also give the $DD(hh, HH, H^+ H^-)$ interaction needed
to evaluate the darkon annihilation rate if the mass of the DM is larger than the $W$, $Z$
and the physical Higgs boson masses~\cite{thdm}.
\begin{eqnarray}
{\cal L}_{VV\cal H} &=&
\Biggl(\frac{2m_W^2}{v}\,W^{+\mu}W_\mu^-+\frac{m_Z^2}{v}\,Z^\mu Z_\mu^{}\Biggr)
\bigl(H\,\sin(\beta-\alpha)+h\, \cos(\beta-\alpha)\bigr)  \,\,,\nonumber\\
{\cal L}_{D\cal H}&=&{1\over 2 \cos\beta } (\lambda_h \cos\alpha - \lambda_H \sin\alpha)
 \left( (\cos\alpha H - \sin\alpha h)^2 + \sin^2\beta A^2 + 2 \sin^2\beta H^+ H^-\right)
 \nonumber\\
&+& {1\over 2 \sin\beta } (\lambda_h \sin\alpha + \lambda_H \cos\alpha)
\left( (\sin\alpha H + \cos\alpha h)^2 + \cos^2\beta A^2 + 2 \cos^2\beta H^+ H^-\right)\;.
\end{eqnarray}

The existence of an additional CP even Higgs boson will modify the SM+D model in several ways.
First of all, the couplings of two Higgs bosons to the SM particles are different not just
from those in the SM+D model but also from each other.
Second, the couplings are able to be adjusted to accommodate a dark matter with mass ranged
from ${\cal O}(1)$ GeV to ${\cal O}(10)$ GeV whose relic density is consistent with
direct DM searches. Finally the additional Higgs boson also provides the possibility of
allowing the lighter one of the two CP even Higgs boson with a small invisible branching
ratio to be discovered at the LHC
while the heavier Higgs boson is responsible for the DM relic density and direct searches.

\section{DM Relic Density, Direct Search and Higgs Width in THDM\,II+D}
Since in general both of the two CP even Higgs bosons $h$ and $H$ couple
to dakron fields, the DM relic density calculation is modified.
The annihilation ratio is given by,
\begin{eqnarray} \label{csan1}
\sigma_{\rm ann}^{}\, v_{\rm rel}^{} \,\,&=&\,\,
\frac{8\lambda_h^2 v^2}{\bigl(4m_D^2-m_h^2\bigr)^2+\Gamma^2_h\,m^2_h}\,
\frac{\sum_i\Gamma\bigl(\tilde h\to X_i^{}\bigr)}{2m_D^{}}\nonumber\\
& +& \frac{8\lambda_H^2 v^2}{\bigl(4m_D^2-m_H^2\bigr)^2+\Gamma^2_H\,m^2_H}\,
\frac{\sum_i\Gamma\bigl(\tilde H\to X_i^{}\bigr)}{2m_D^{}} \,\,,
\end{eqnarray}
where $\Gamma\bigl(\tilde H \to X_i^{}\bigr)$ indicates the decay width
 of $H$ into SM particles with a virtual mass of $2 m_D$, similar to
that for $h$ in Eq.~(\ref{csan}).

The cross-section of the darkon-nucleon
elastic scattering is also modified to include two contributions
\begin{eqnarray} \label{cs_el_2hdm}
\sigma_{\rm el}^{} \,\,\simeq\,\,
\frac{m_N^2\,v^2}{\pi\bigl(m_D^{}+m_N^{}\bigr)^2} \Biggl(\frac{\lambda_h^{}\,g_{NNh}^{\rm THDM}}{m_h^2}
+ \frac{\lambda_H^{}\,g_{NNH}^{\rm THDM}}{m_H^2}\Biggr)^{\!2} \,\,,
\end{eqnarray}
where, from Eq.~(\ref{gnnh0}) and results in Ref.~\cite{sm+d},
the nucleon coupling to \,${\cal H}=h$ or $H$\, is
\begin{eqnarray} \label{gnnh_2hdm}
g_{NN\cal H}^{\rm THDM} \,\,=\,\,
\bigl(k_u^{\cal H}-k_d^{\cal H}\bigr) \frac{\sigma_{\pi N}^{}}{2 v} \,+\,
k_d^{\cal H}\, \frac{m_N^{}}{v}
\,+\, \frac{4k_u^{\cal H}-25k_d^{\cal H}}{27}\,\, \frac{m_B^{}}{v} \,\,.
\end{eqnarray}
The parameters $k_q^{\cal H}$ are read off from Eq.~(\ref{yukawa}) to be
\begin{eqnarray}  \label{ksm'}
k_u^h  \,\,=\,\, \frac{\sin\alpha}{\sin\beta} \,\,, \hspace{2em}
k_d^h \,\,=\,\,  \frac{\cos\alpha}{\cos\beta} \,\,, \hspace{1cm}
k_u^H  \,\,=\,\, \frac{\cos\alpha}{\sin\beta} \,\,, \hspace{2em}
k_d^H \,\,=\,\, -\frac{\sin\alpha}{\cos\beta} \,\,.
\end{eqnarray}

Due to the extra new parameters in the model, there are more possibilities compared with the SM+D model
in regard to the range of DM mass and the branching ratio of the Higgs boson invisible decay.

The most similar way to SM+D is that
the lighter one of the two CP-even Higgs bosons, $h$ or $H$, will be discovered
at the LHC and this Higgs boson also couples to darkon pairs responsible to produce
the right amount of DM relic density.
Similar to the results found in Ref.~\cite{sm+d} that without cancelation among
different contributions to the Higgs boson interaction with nucleon,
the DM mass roughly in the range of 10 GeV to a half of the Higgs boson mass is ruled out,
approximately leaving a small range of 2 GeV to 10 GeV allowed.

If the DM mass turns out to be roughly between 10 GeV and a half of the Higgs boson mass,
additional cancelation mechanism should be in effect to accommodate this.
It has been pointed out that this cancelation mechanism can indeed happen~\cite{sm+d}.
Because of the extra new parameters in the model we consider,
cancelation among different contributions to the DM direct search is possible
if
\begin{eqnarray}
\frac{k_d^{\cal H}}{k_u^{\cal H}} \,\,=\,\,
\frac{27\,\sigma_{\pi N}^{}+8\,m_B^{}}{27\,\sigma_{\pi N}^{}+50\,m_B^{}-54\,m_N^{}} \,\,.
\end{eqnarray}
Since $m_B^{}$ is related to $\sigma_{\pi N}^{}$ and $\sigma_{\pi N}^{}$
is not well determined, the numerical value of $k_d^h/k_u^h$ has a sizable
uncertainty for $35{\rm\,MeV}\lesssim\sigma_{\pi N}^{}\lesssim80$\,MeV\,
~\cite{Cheng:1988cz,gasser,Ellis:2008hf}.
Nevertheless, we have checked that the main conclusion of this section
still holds for $\sigma_{\pi N}^{}$ in this range.
For definiteness, from now on we employ  \,$\sigma_{\pi N}^{}=45$\,MeV\,~\cite{Cheng:1988cz}.

Given these input values,
there is a cancelation in the direct DM search cross section,
if numerically $k^{\cal H}_d/k^{\cal H}_u  = -0.405$.
Models with $\alpha$ and $\beta$ which keep $k^{\cal H}_d/k^{\cal H}_u$ in the vicinity of $-0.405$
will then have small direct DM search cross section allowing the DM mass to be
approximately in the range of 10 GeV to a half of the Higgs boson mass.
This is different from that in SM+D model.
If the numerical values of $k_d^{\cal H}/k_u^{\cal H}$ are not exactly the critical value
so that the cancelation in direct detection cross section is not complete,
it then leaves some chances for direct DM search. We will choose \,$k_d^{\cal H}/k_u^{\cal H} = -0.42$ for discussions.
Future experimental searches can narrow down the parameter space~\cite{superCDMS,xenon100p}.

\begin{figure}[htp]
\includegraphics[width=0.45\textwidth]{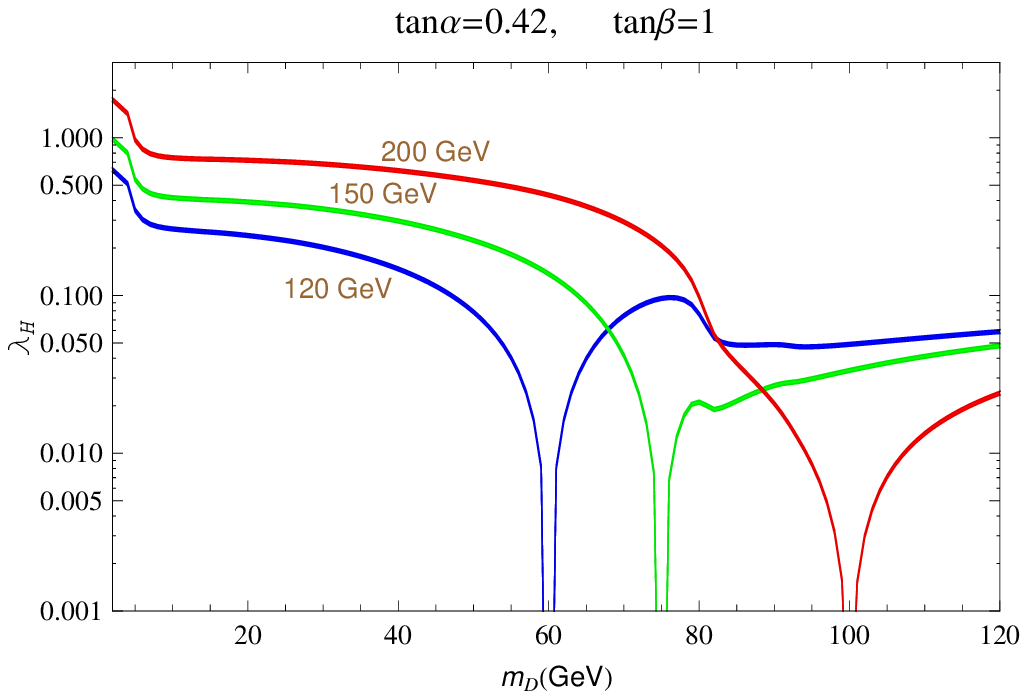}
\includegraphics[width=0.45\textwidth]{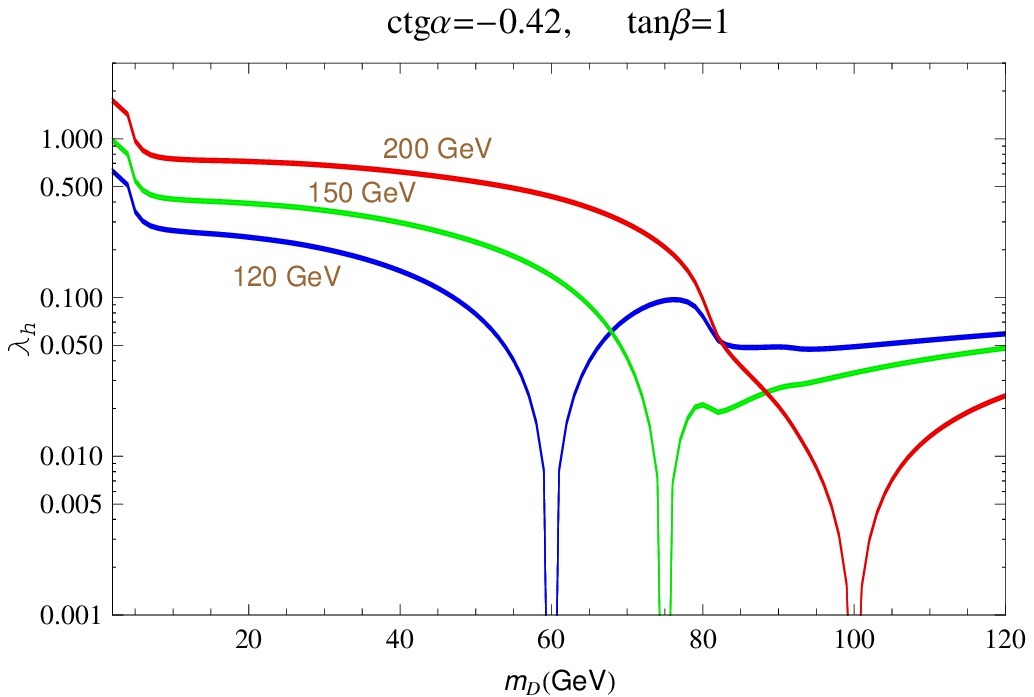}\\
\caption{Darkon-Higgs couplings $\lambda_H$ (left) and $\lambda_h$ (right) as  functions of darkon mass $m_D$
with different Higgs masses $m_H$ (left) and $m_h$ (right) with $\tan\beta =1$ in THDM+D. \label{fig:2hdmd}
}
\end{figure}

\begin{figure}[htp]
\includegraphics[width=0.45\textwidth]{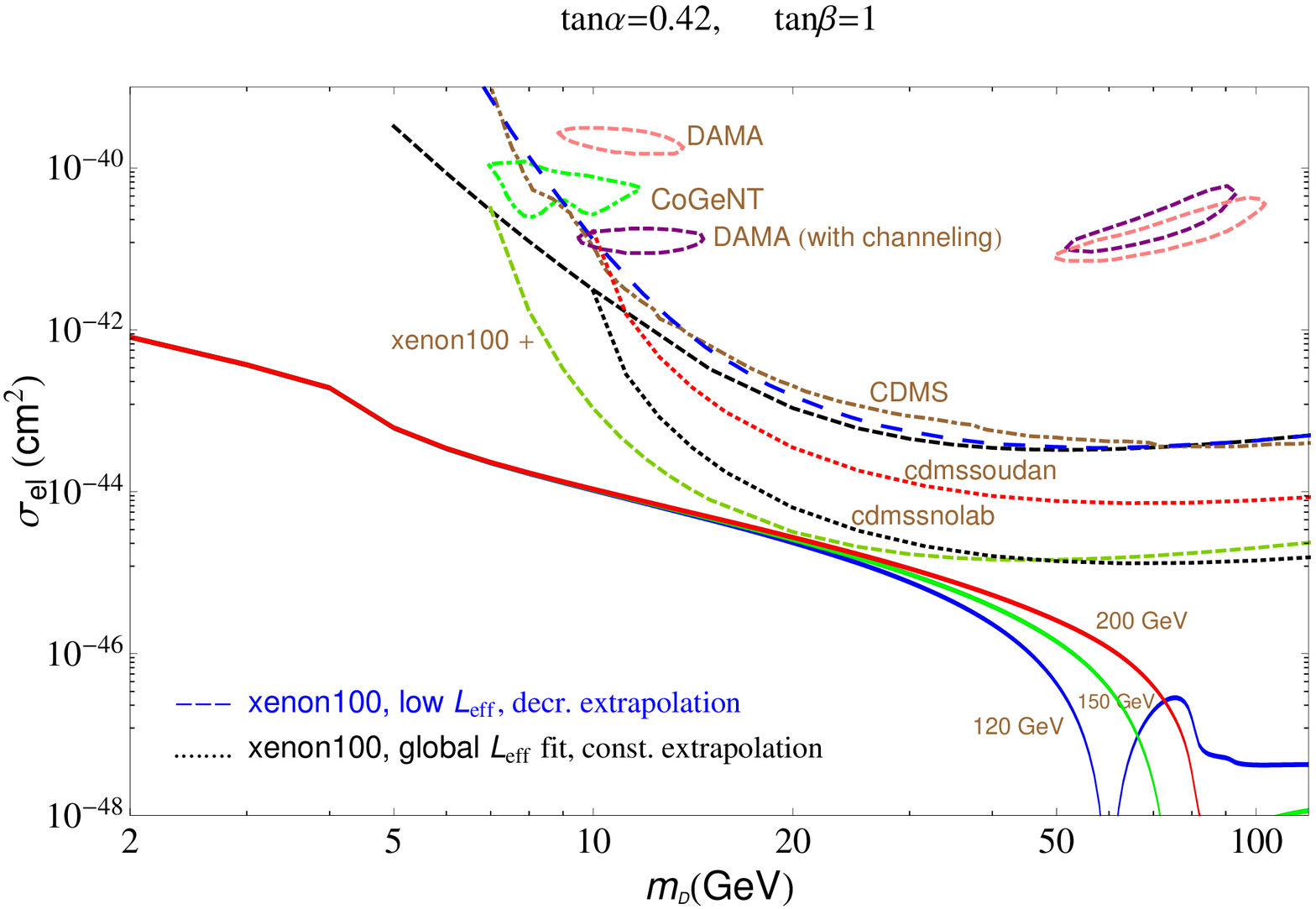}
\includegraphics[width=0.45\textwidth]{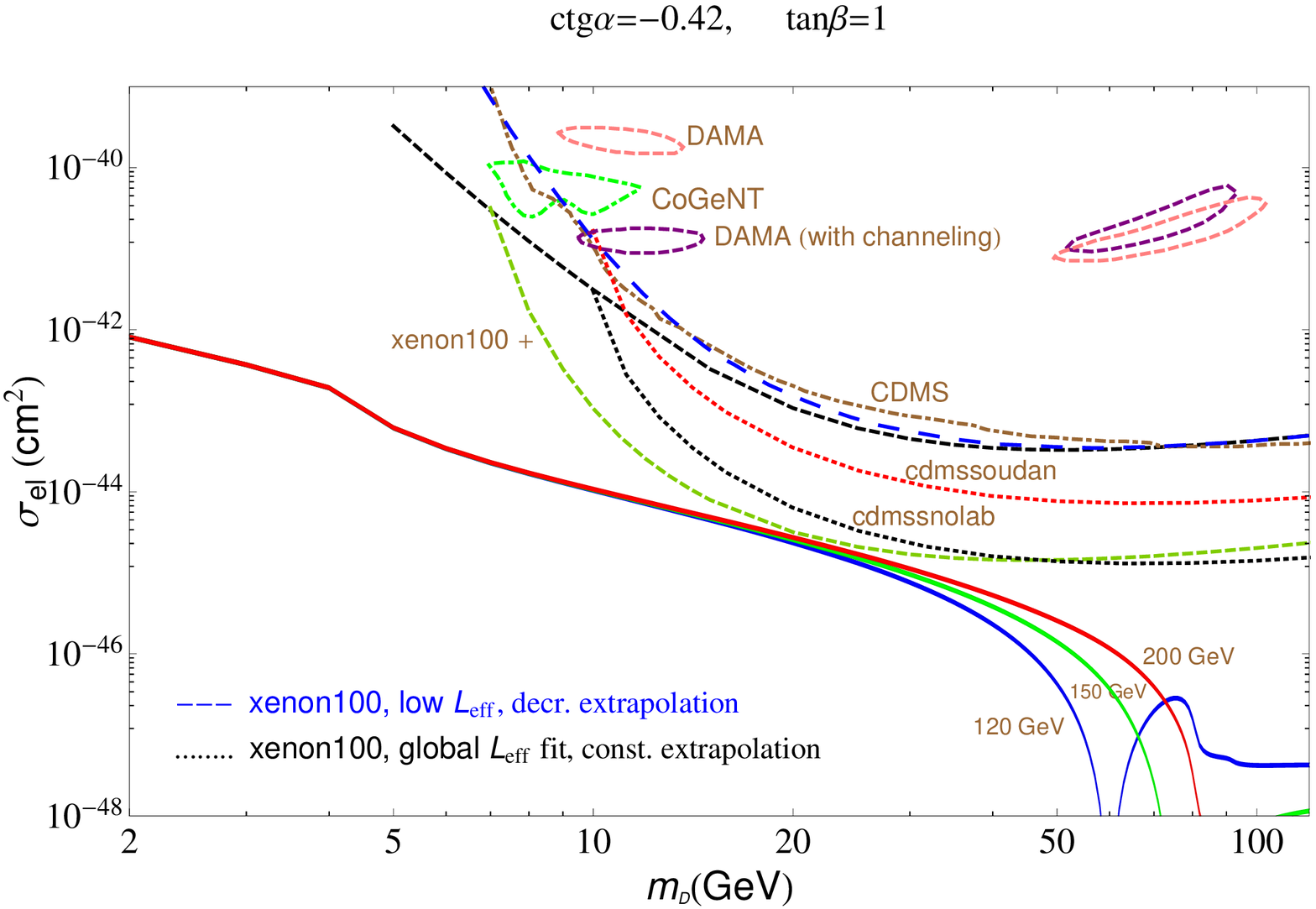}\\
\caption{Darkon-Nucleon elastic cross sections $\sigma_{\rm el}$ as  functions of darkon mass $m_D$
with different Higgs masses $m_H$ (left) and $m_h$ (right) with $\tan\beta = 1$ in THDM+D, compared to 90\% C.L. upper
limits from experimental data. \label{fig:2hdmdsigma}
}
\end{figure}

\begin{figure}[htp]
\includegraphics[width=0.45\textwidth]{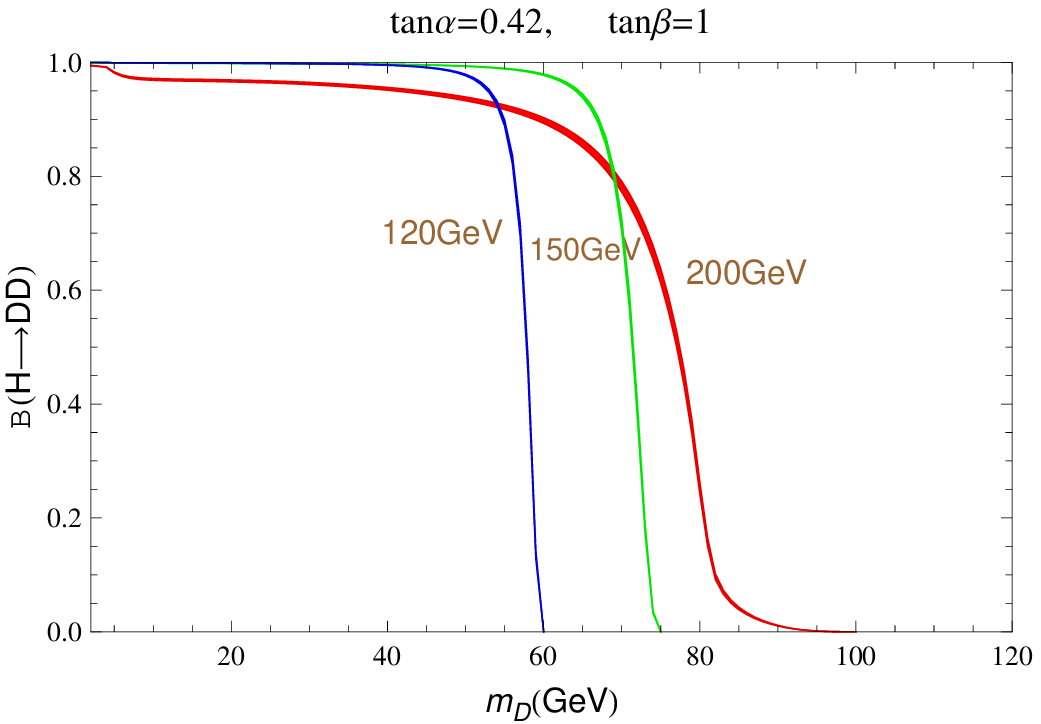}
\includegraphics[width=0.45\textwidth]{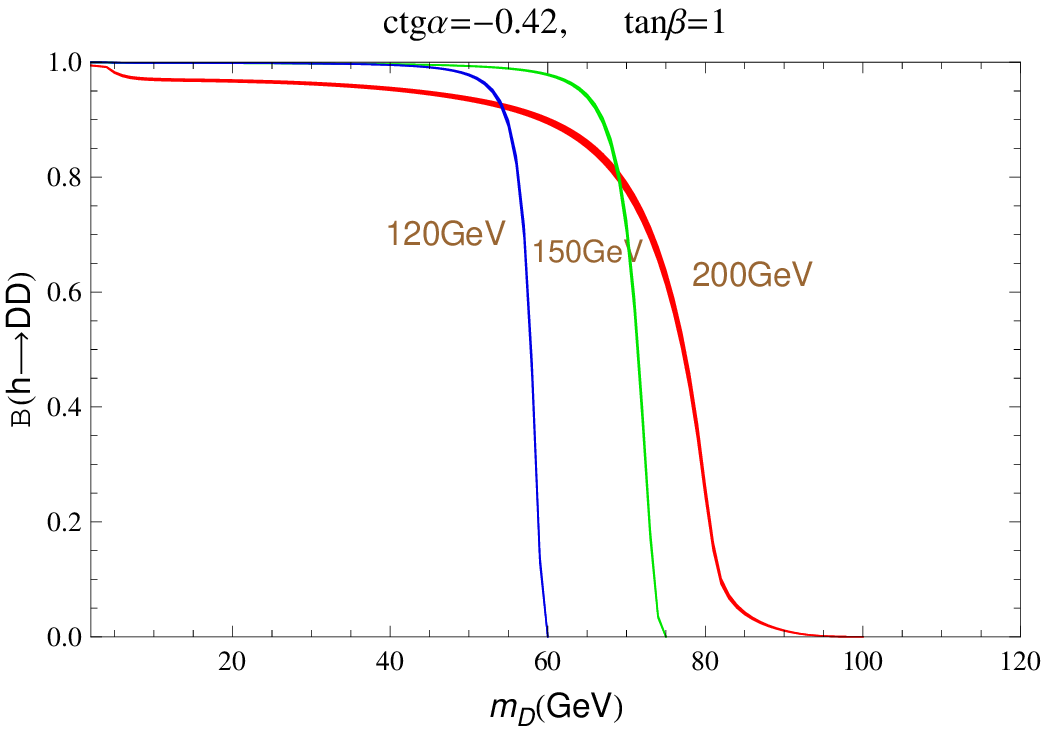}\\
\caption{
The invisible decay branching ratios of the Higgs boson
as functions of darkon masses $m_D$ with different Higgs boson mass $m_H$ (left) and $m_h$ (right)
with $\tan\beta = 1$ in THDM+D. \label{fig:2hdmdbr}}
\end{figure}

For this case, we find that even a smaller invisible Higgs decay branching ratio can be made by changing the value of $\tan\beta$
since this can lead to a smaller $\lambda_{h,H}$ compared to $\lambda$ in SM+D,
it is still significantly larger than the SM prediction.
If the LHC will find a light Higgs with very small invisible Higgs boson
branching ratio, this scenario will be ruled out.
In Figs. \ref{fig:2hdmd}, \ref{fig:2hdmdsigma} and \ref{fig:2hdmdbr} we show two examples of this scenario
assuming that the lighter Higgs boson is $H$ or $h$ which is also responsible to
DM physics.  In the case for $H$,  $k_d^{\cal H}/k_u^{\cal H} = k_d^H/k_u^H = - \tan\alpha \tan\beta$,
and in the case for $h$,
$k_d^{\cal H}/k_u^{\cal H} = k_d^h/k_u^h = \tan\beta/\tan\alpha$. In both cases we use $\tan\beta = 1$
with $m_H = 120, \, 150, \, 200 \, $ GeV for illustrations.
There are some detailed differences for the two cases, but the general features are similar. It can be seen from
the plots that the Higgs invisible branching ratios are very large. We have checked for other values of $\tan\beta$.
We find that although by varying $\tan\beta$, a smaller invisible width
for $H(h) \to DD$ can be obtained, the invisible branching ratio is always very large.

%%%%%%%%%%%%%%%%%%%%%%%%
Given two CP-even Higgs bosons, it is also possible that the lighter Higgs boson
which will be discovered at the LHC does not play a significant role in the DM physics,
and the DM relic density is mainly determined by the interaction between the heavier Higgs
and the darkon. This possibility can be easily realized if the coupling constant of the lighter Higgs to be discovered at the LHC
and the darkon pair, $\lambda_{h,H}$,
is almost zero such that it is not directly related DM annihilation and  direct search.

We now discuss these possibilities in detail.
We will consider two cases,
a) $h$ is lighter than $H$ with $\lambda_h = 0$,
and b) $H$ is lighter than $h$ with $\lambda_H = 0$.
%%% case a

In case a), since $\lambda_h = 0$, even if $m_D$ is smaller than $m_h/2$,
the process $h \to DD$ has vanishing decay width and therefore
a large invisible branching ratio of $h$ is forbidden.
The branching ratios of $h$ for other decay modes, however,
will be different than those in SM+D
because of the dependence on the mixing angles $\alpha$ and $\beta$.
If the coupling $\lambda_H$ is not zero, $H$ will interact with darkon.
Requiring this interaction to produce the right DM relic density,
the parameters are constrained.
The same interaction will control the direct DM search cross section.
One can use the same cancelation mechanism for direct DM search cross section discussed earlier to allow DM mass as wide a range as possible.

\begin{figure}[htp]
{\label{fig:2hdmATan1lambda}\includegraphics[width=0.45\textwidth]{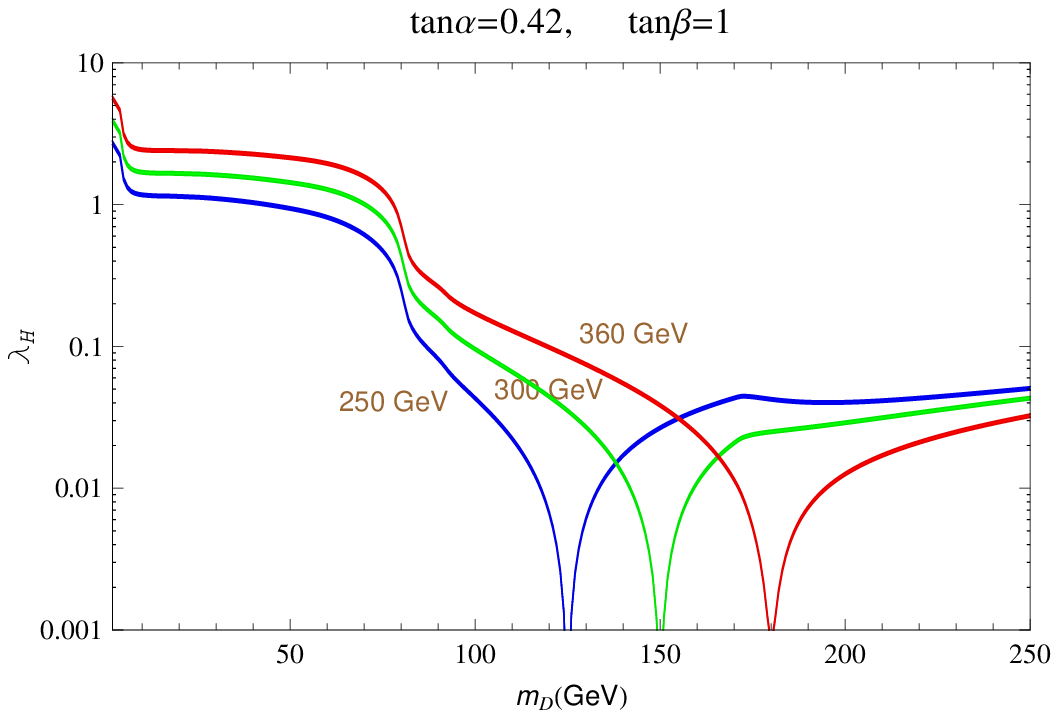}}
{\label{fig:2hdmATan30lambda}\includegraphics[width=0.45\textwidth]{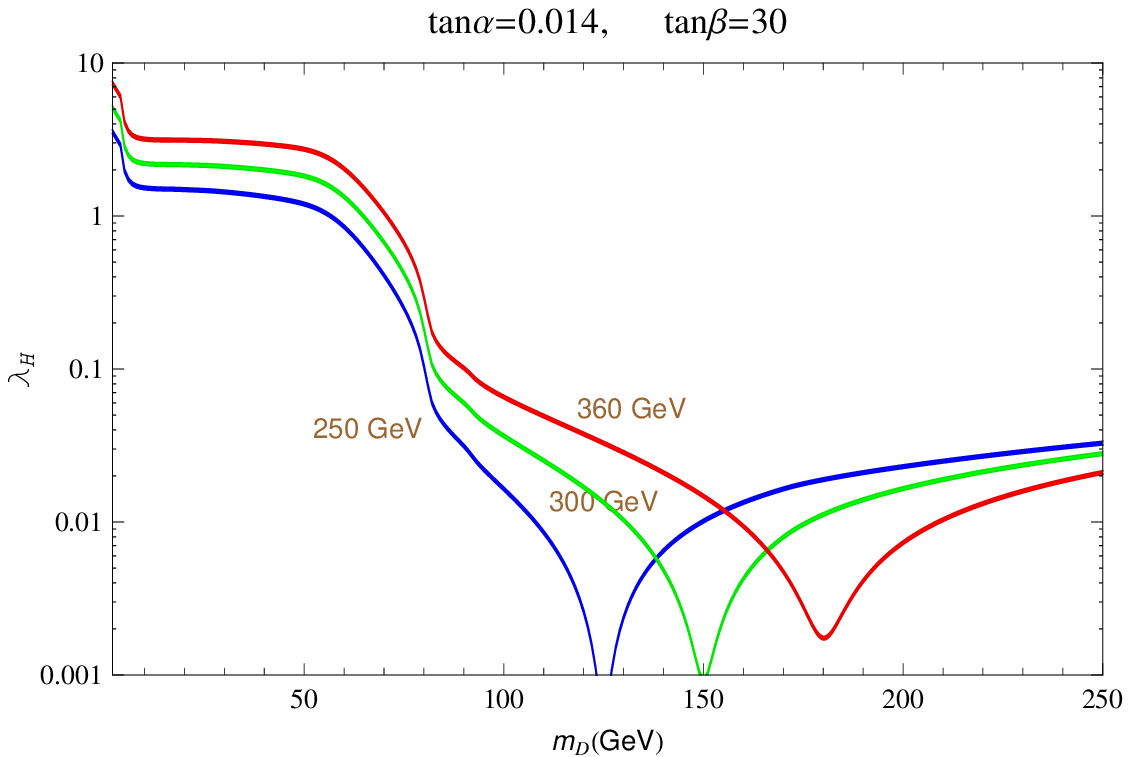}}
\caption{Darkon-Higgs coupling $\lambda_H$ as a function of darkon mass $m_D$
 with different Higgs masses $m_H$ and $\tan\beta$ in THDM+D. \label{2HDMcasea-lambda}}
 \end{figure}

\begin{figure}[htp]
\includegraphics[width=0.45\textwidth]{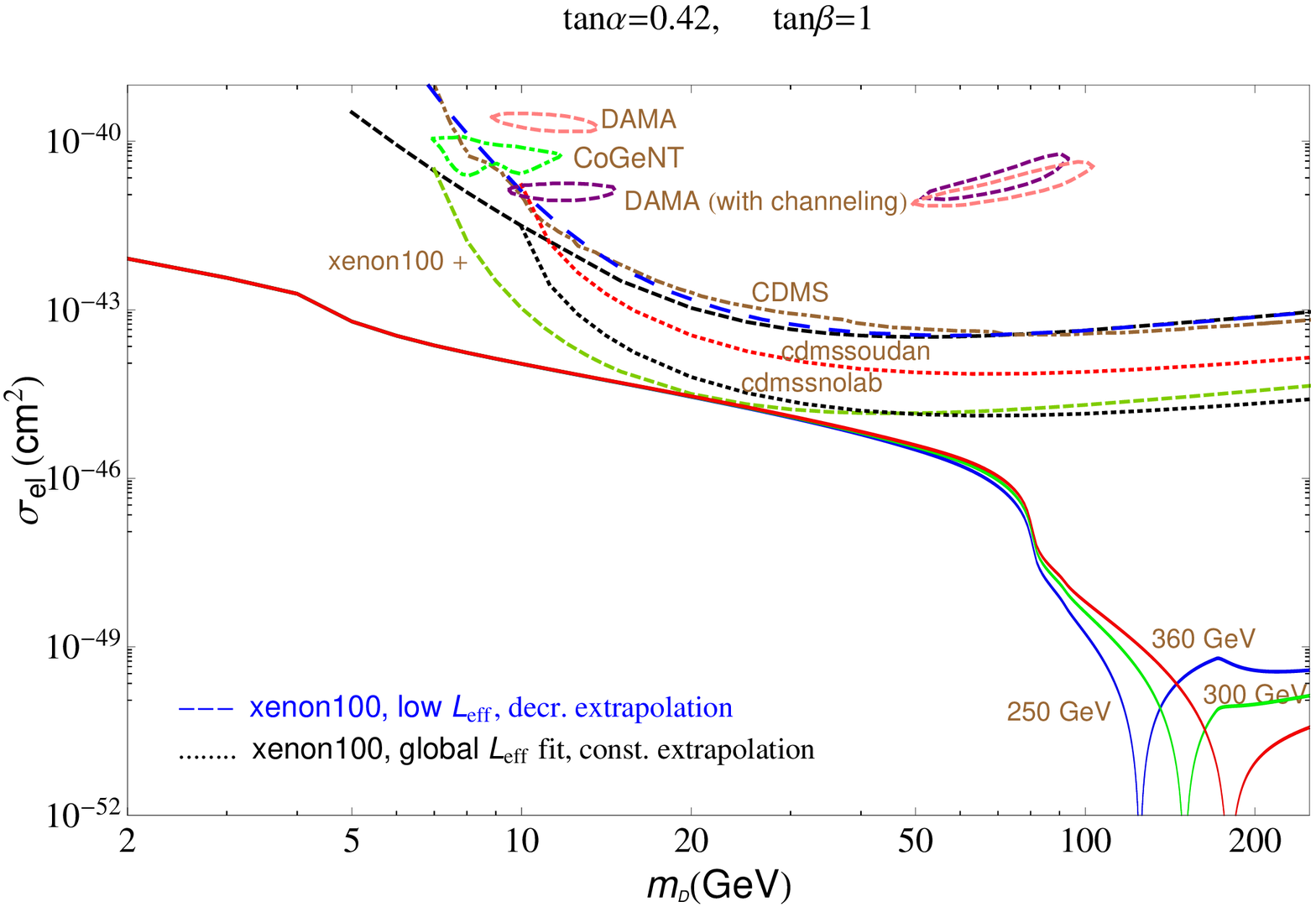}
\includegraphics[width=0.45\textwidth]{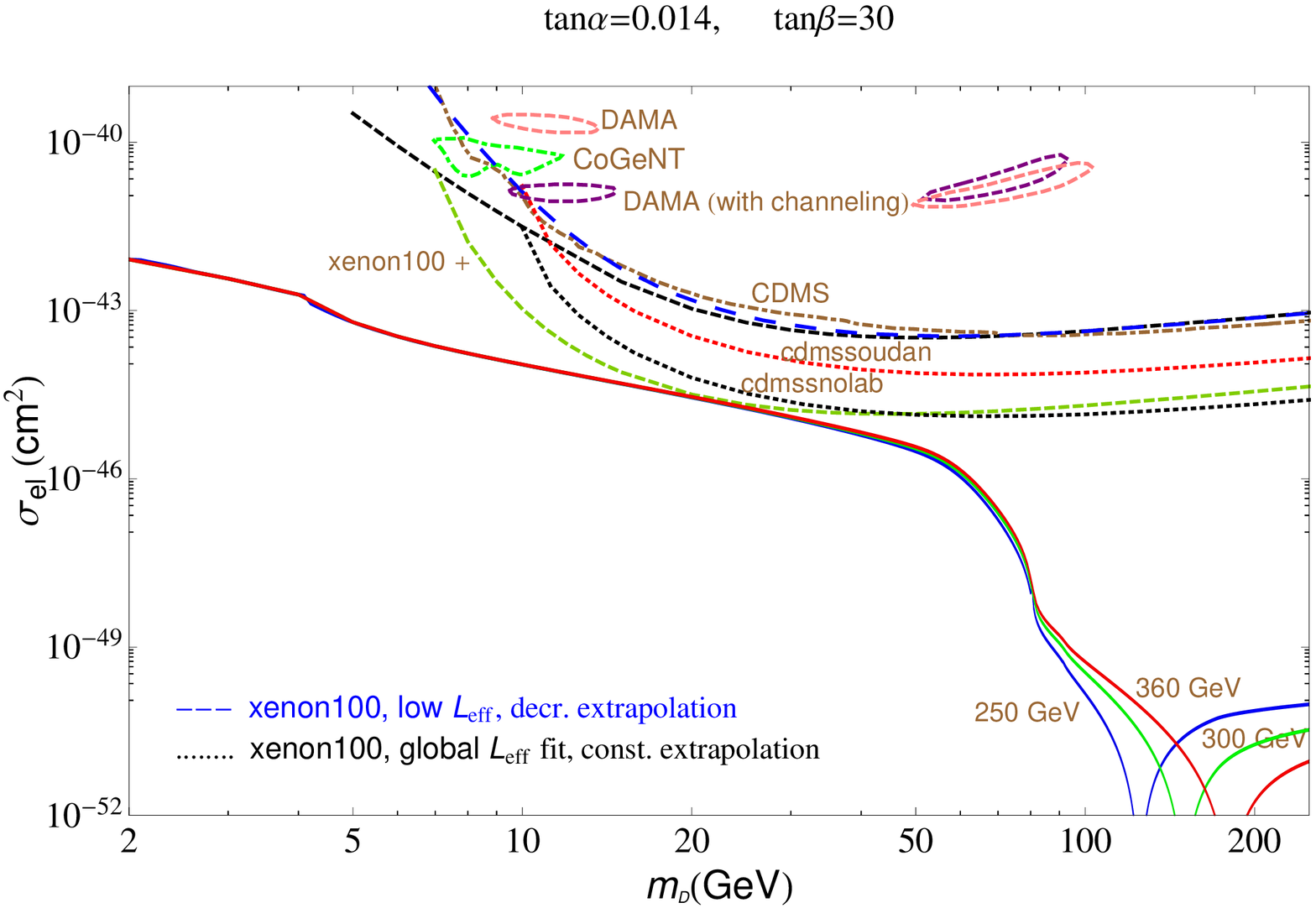}
\caption{Darkon-Nucleon elastic cross section $\sigma_{\rm el}$ as a function of darkon mass $m_D$
with different Higgs masses $m_H$ and $\tan\beta$ in THDM+D, compared to 90\% C.L. upper
limits from experimental data. \label{2HDMcasea-sigma}}
\end{figure}

\begin{figure}[htp]
{\label{fig:2hdmATan1br}\includegraphics[width=0.45\textwidth]{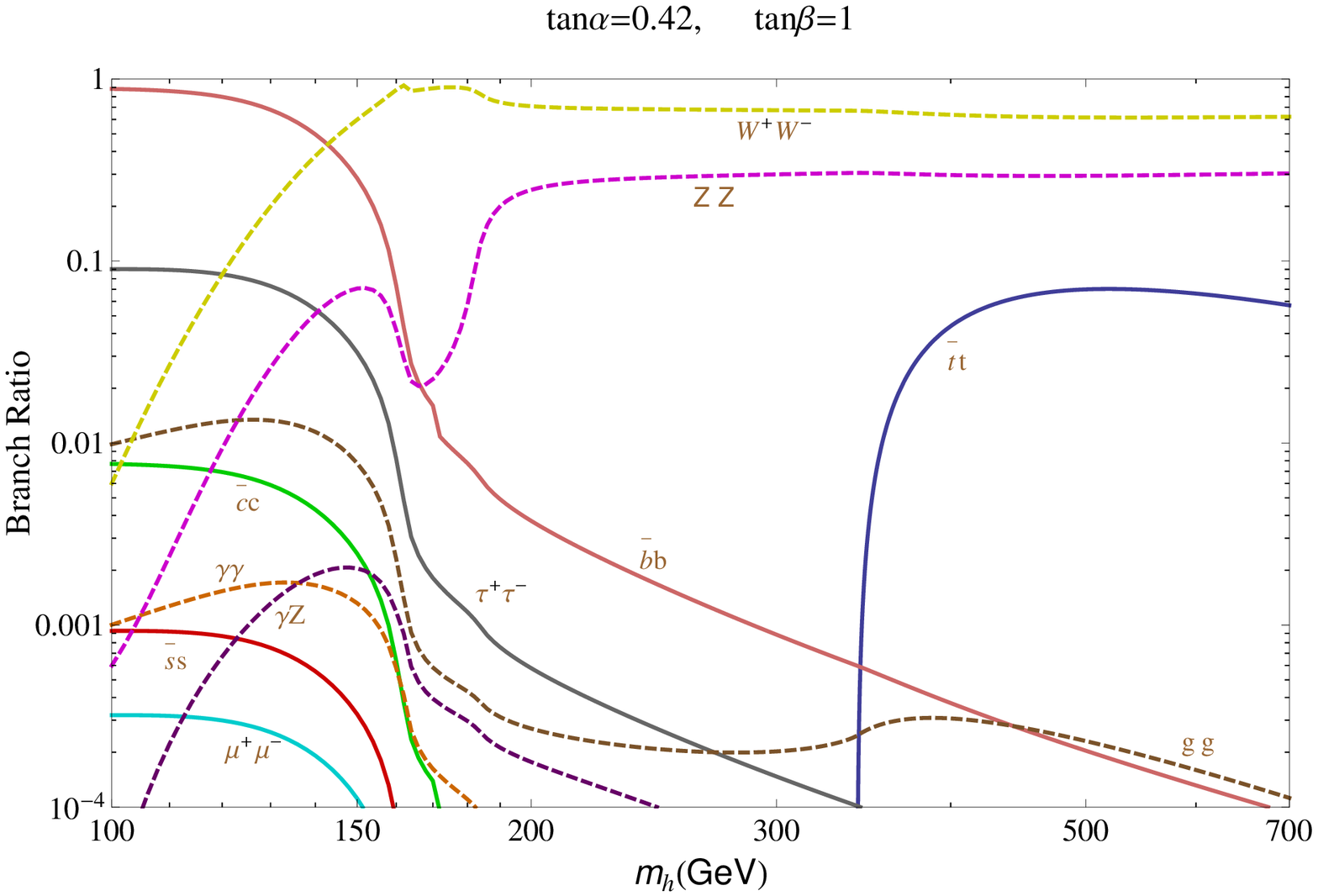}}
{\label{fig:2hdmATan30br}\includegraphics[width=0.45\textwidth]{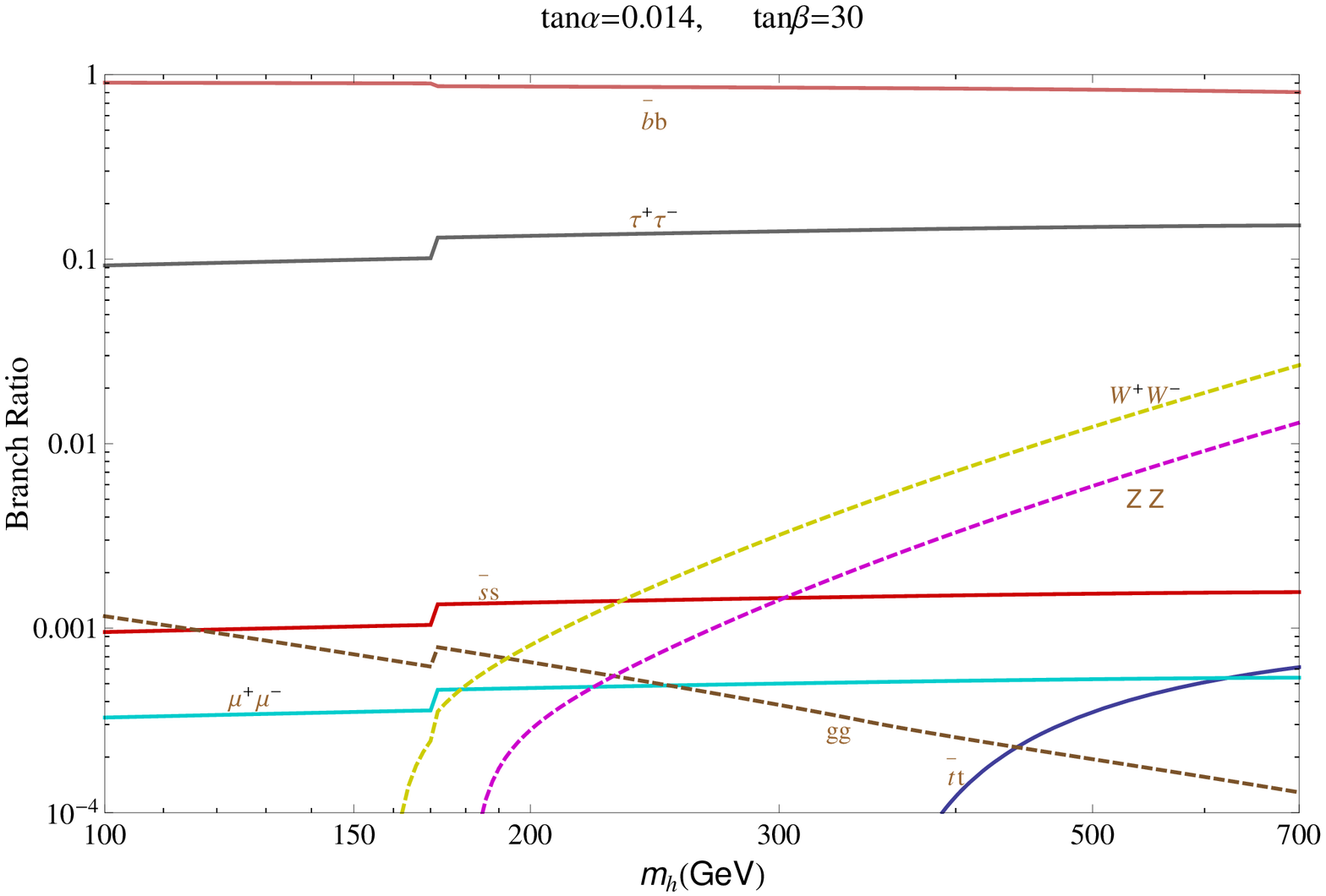}}
\caption{The branching ratio of the lighter Higgs boson
as a function of lighter Higgs boson mass $m_h$ in THDM+D with different values of $\tan\beta$.\label{2HDMcasea-br}}
\end{figure}

We show the results in Figs.~\ref{2HDMcasea-lambda}, \ref{2HDMcasea-sigma} and \ref{2HDMcasea-br}, where
the numerical values of $k_d^H/k_u^H$ are chosen, again, not exactly the critical value
so that the cancelation in direct detection cross section is not complete with $k_d^H/k_u^H = -\tan\alpha\tan\beta= -0.42$.
Future experimental searches can narrow down the parameter space.
We consider two sets of $\tan\alpha$ and $\tan\beta$ values satisfying this choice:
\,$(\tan\alpha,\tan\beta)=(0.42,1)$\,
and \,$(\tan\alpha,\tan\beta)=(0.42/30,30)$
with $m_H = 250, \, 300, \, 360 \, $ GeV for illustrations.
Since small $\tan\beta$ is disfavored for low mass charged-Higgs $H^\pm$~\cite{WahabElKaffas:2007xd},
we will assume the charged-Higgs has a mass larger than 250 GeV.
We notice that the two choices of $\tan\beta$ roughly span the range
allowed by various experimental and theoretical constraints~\cite{WahabElKaffas:2007xd}.

In Fig.~\ref{2HDMcasea-lambda}
the coupling constant $\lambda_H$ is bigger than 1 in
the low $m_D$ range which seems to spoil the perturbation.
But a careful study of the perturbative unitarity of darkon-Higgs interaction
at the tree level~\cite{Cynolter:2004cq} suggests that perturbation would not break down
if $|\lambda_H| < 4\pi$.  The obscurity between perturbation and non-perturbation~\cite{pt}
also suggests a less constrained condition $|\lambda_H| < 2\sqrt{\pi} (m_H /100 GeV)^2$.
In all, the perturbativity of the theory is not spoiled in the parameter space we
choose.

%%%%% case b

\begin{figure}[htp]
{\label{fig:2hdmBTan1lambda}\includegraphics[width=0.45\textwidth]{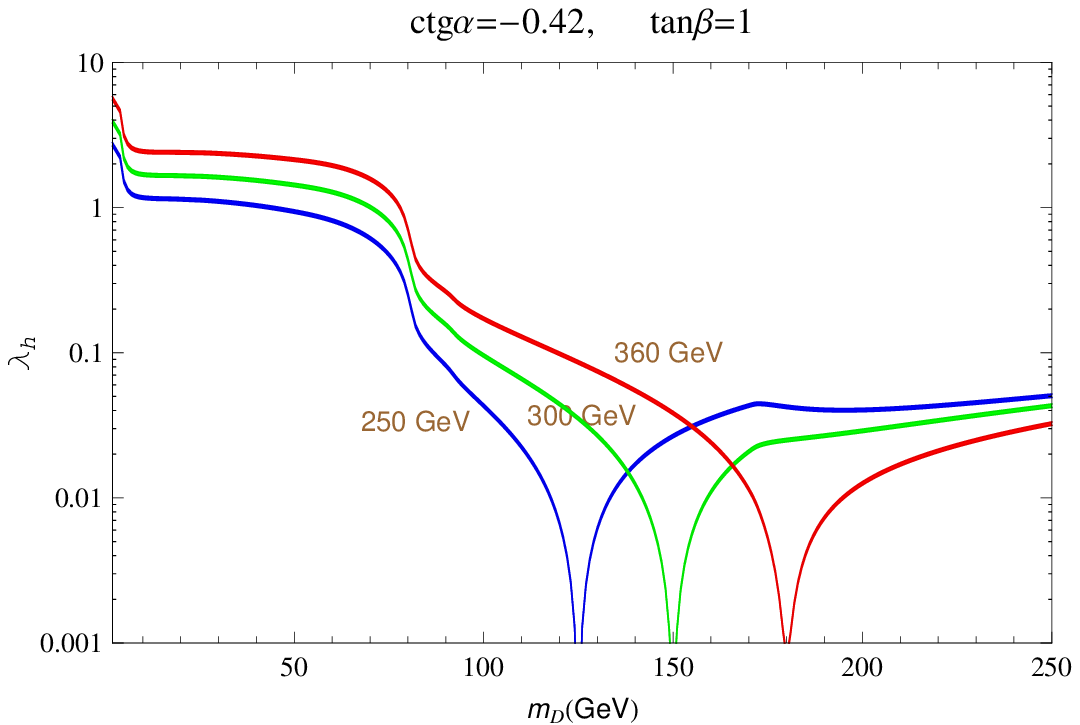}}
{\label{fig:2hdmBTan30lambda}\includegraphics[width=0.45\textwidth]{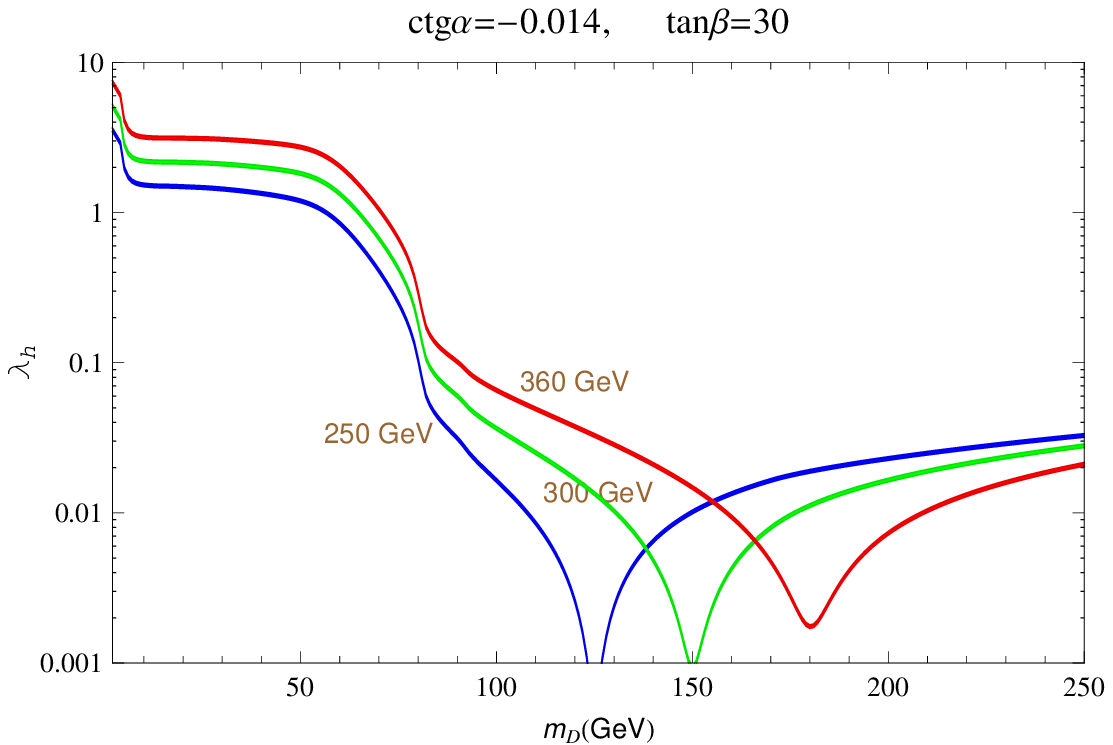}}
\caption{Darkon-Higgs coupling $\lambda_h$ as a function of darkon mass $m_D$
 with different Higgs masses $m_h$ and $\tan\beta$ in THDM+D. \label{2HDMcaseb-lambda}}
\end{figure}

\begin{figure}[htp]
{\label{fig:2hdmBTan1sigma}\includegraphics[width=0.45\textwidth]{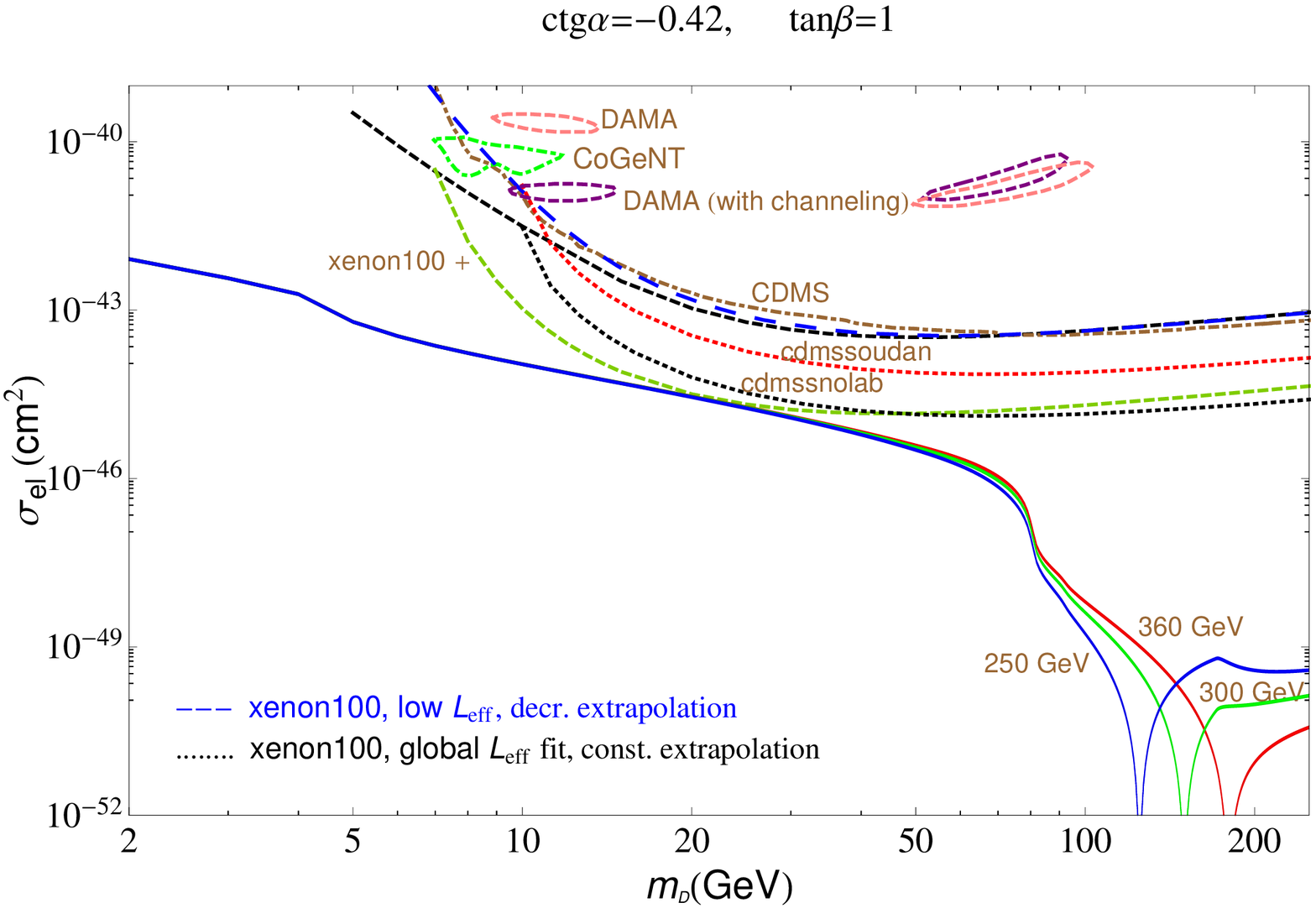}}
{\label{fig:2hdmBTan30sigma}\includegraphics[width=0.45\textwidth]{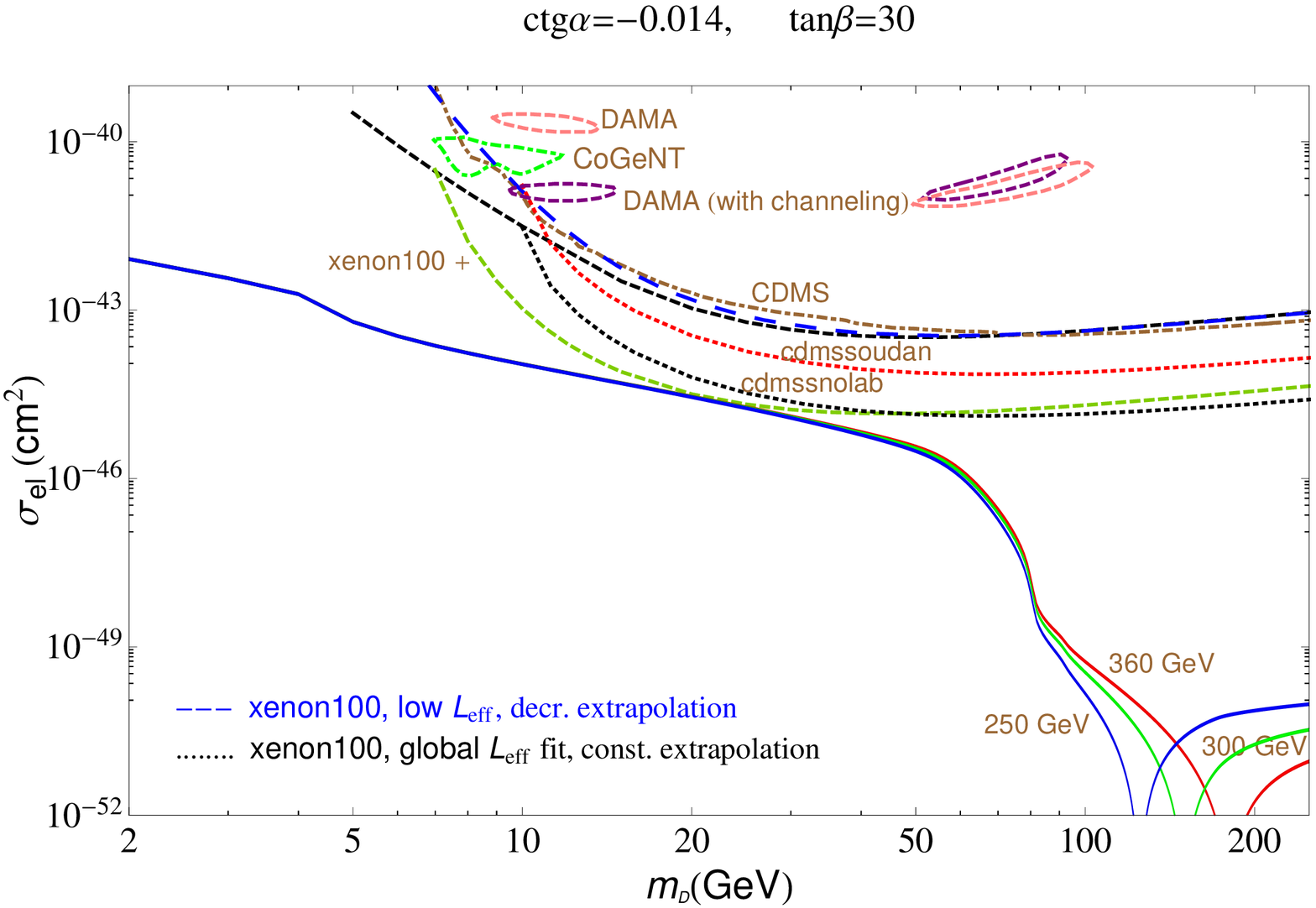}}
\caption{Darkon-Nucleon elastic cross section $\sigma_{\rm el}$ as a function of darkon mass $m_D$
with different Higgs masses $m_h$ and $\tan\beta$ in THDM+D, compared to 90\% C.L. upper
limits from experimental data. \label{2HDMcaseb-sigma}}
\end{figure}

\begin{figure}[htp]
{\label{fig:2hdmBTan1br}\includegraphics[width=0.45\textwidth]{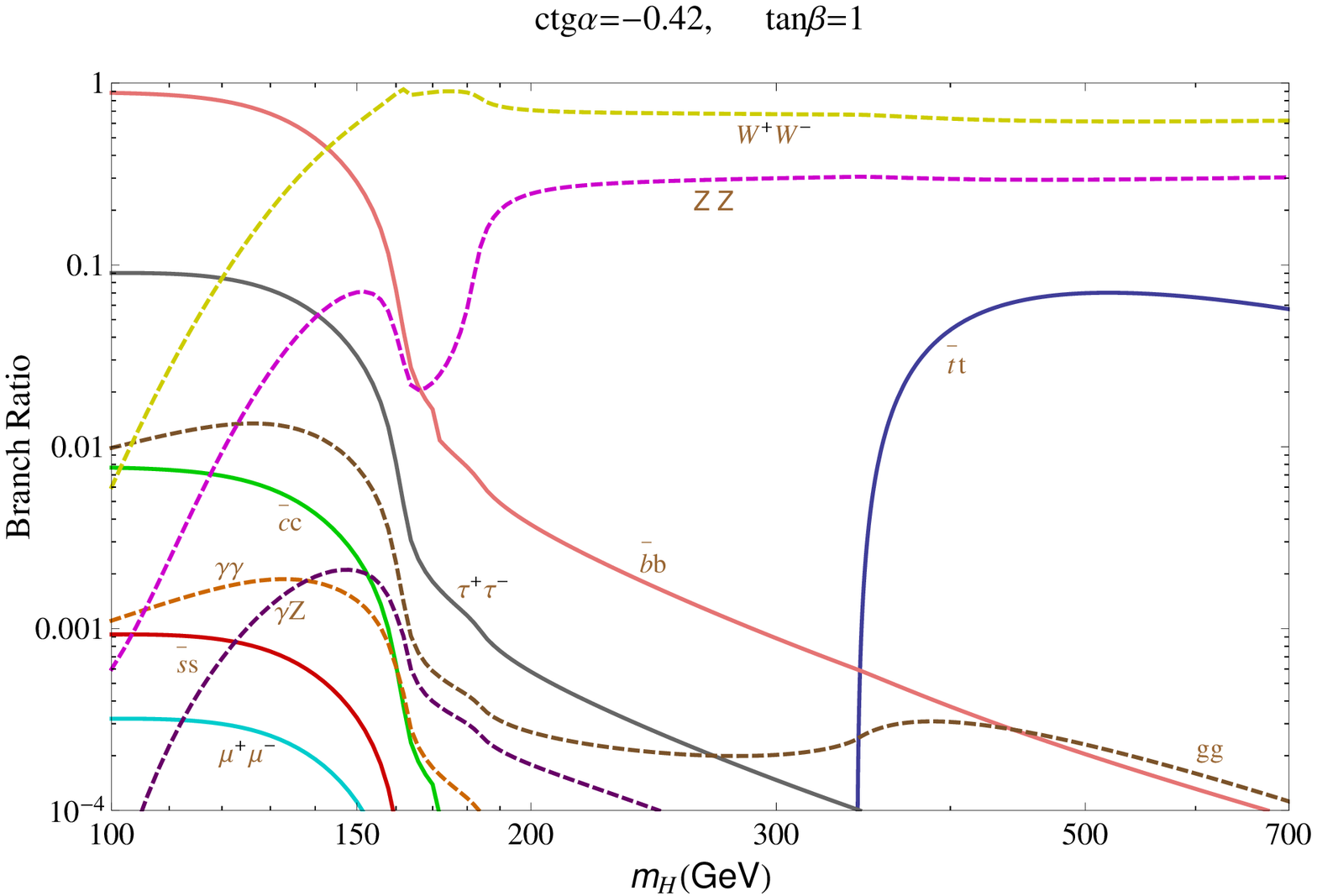}}
{\label{fig:2hdmBTan30br}\includegraphics[width=0.45\textwidth]{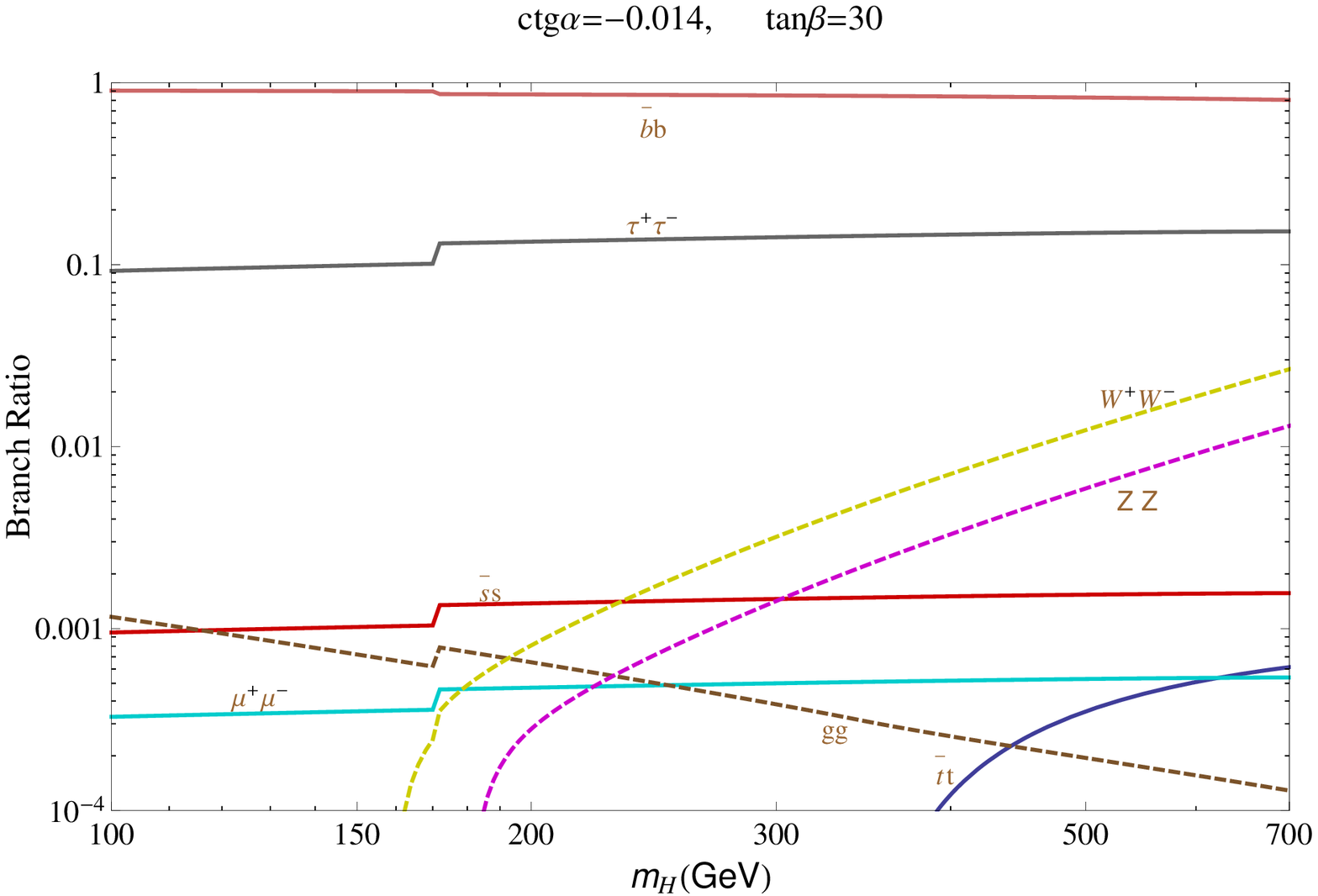}}
\caption{The branching ratio of the lighter Higgs boson
as a function of the lighter Higgs boson mass $m_H$ in THDM+D with different values of $\tan\beta$. \label{2HDMcaseb-br}}
\end{figure}

In case b), $h$ and $H$ just exchange their roles.
Thus the condition for the cancelation in direct detection cross section will be modified to,
$k_d^{\cal H}/ k_u^{\cal H} = k_d^h/ k_u^h = \tan \beta/\tan \alpha$.
We choose  again \,$k_d^{\cal H}/k_u^{\cal H}= -0.42$\,
and consider two sets of $\tan\alpha$ and $\tan\beta$ values satisfying this choice:
\,$(\tan\alpha,\tan\beta)=(-1/0.42,1)$\,
and \,$(\tan\alpha,\tan\beta)=(-30/0.42,30)$
with $m_H = 250, \, 300, \, 360 \, $ GeV for illustration.
The results  are shown in Figs. \ref{2HDMcaseb-lambda}, \ref{2HDMcaseb-sigma} and \ref{2HDMcaseb-br}.

In both cases, a) and b), the allowed parameters can
make the direct detection cross section to be close to the current bounds.
Future experimental data can further narrow down the parameter spaces.
The difference with different choices of $\tan\beta$ is more prominent
in the lighter Higgs boson decay branching ratios.
With small and large $\tan\beta$, the lighter Higgs boson branching ratios arevery different.
This is because for large $\tan\beta$ the branching ratios of
Higgs to down-type fermion decay are largely enhanced due to large down-type Yukawa
couplings with large $\tan\beta$. Comparing experimental data with theoretical predictions for branching ratios, 
information on mixing parameter $\alpha$ of the two CP even Higgs boson can be obtained. Therefore LHC 
data can also help to further narrow down the parameter space.

\section{Discussions and Conclusions}
As the simplest extension to the Standard Model, the SM+D not only provides
a dark matter candidate but also modifies the Higgs physics dramatically.
This model has only a smaller number of parameters.
A big range of DM mass, roughly from 10 GeV to half of the
Higgs boson mass, is ruled out from experimental data on DM relic density
and direct search cross section.
The Higgs boson in this model always has very large invisible branching ratio
if the darkon is lighter than a half of the
Higgs boson mass, because of the
sizable darkon-Higgs interaction coupling. As seen in Fig.~\ref{fig:smdbr},
the Higgs boson invisible branching ratio is increasing for smaller Higgs mass, since the Higgs
partial width into SM particles is suppressed.
The large invisible widths will change the traditional paradigm used for Higgs hunting.
Although difficult,
it may still be possible for such a Higgs to be detected directly at CMS through the usual SM modes
with 30 fb$^{-1}$ of integrated luminosity~\cite{barger} or by missing transverse energy
search at ATLAS~\cite{barger, invisiblehiggs3, Davoudiasl:2004aj}.
If no large invisible decay width for Higgs boson will be found, the possibility of a
low DM mass smaller than half of the Higgs boson mass will be ruled out.

In THDM+D, there are two CP-even Higgs bosons and both couple to the DM darkon field.  There are a few extra parameters,
which does not help to make the Higgs invisible width much smaller if we continue to consider
the strategy in SM+D, namely the Higgs boson to be detected at the LHC is also the one
responsible for the DM relic density.  If we change the strategy and let one Higgs boson to be
detected at the LHC and the other one to be responsible for the DM relic density, the situation
changes instantly. It is possible to have cancelation in the direct DM detection cross section allowing DM mass
to be in the range ruled out in the SM + D model.
The decay branching ratio of the Higgs boson, which is the usual one to be detected at the LHC,
shows a familiar change for small and large $\tan\beta$ as in the other two Higgs doublet models like
the THDMII and Minimal Supersymmetric Standard Model.

In conclusion, we have studied the constraints from DM relic density,
direct searches and some implications for the collider experiments in
both SM+D and THDM+D.  In the SM+D,
a darkon with mass roughly in the ranges of 10 GeV
 to half of the Higgs boson mass is ruled out as a WIMP candidate by the direct searches.
For DM roughly lighter than 10 GeV,
the invisible branching ratio is always very large which may be ruled out if a small invisible width for Higgs boson will be found at the LHC.
In THDM+D, the experimental limits can be circumvented due to suppression of the darkon-nucleon elastic
cross-section at some values of $\alpha$ and $\beta$.
However, if the Higgs boson is responsible for dark matter physics and also be detected at the LHC,
using the darkon-Higgs coupling extracted from the DM relic density,
the invisible branching ratio of the would be substantial increase by a large
contribution from the invisible mode \,$h\to DD$, if kinematically allowed.\,
If a Higgs boson with a small invisible decay width will be found at the LHC, the possibility will also be in trouble.
We find that with THDM+D, it is possible to adjust parameters
such that one of the Higgs bosons is primarily responsible for the DM relic density.
The additional Higgs boson to be the lighter one to be detected at the LHC with  a small invisible branching ratio.
Future DM search experiments can further constrain the parameter space of the model.
This could significantly affect Higgs searches at the LHC, we expect that it will still
be able to probe the darkon model.
\\

\acknowledgments \vspace*{-1ex}
This work was partially supported by NSC, NCTS,  NNSF and SJTU Innovation Fund for Postgraduates and Postdocs.
We thank P. Gagnon and J. Tandean for some useful discussions.

\end{document}